# Disentangling the Competing Mechanisms of Light-Induced Anomalous Hall Conductivity in Three-Dimensional Dirac Semimetal

Yuta Murotani[1]*, Natsuki Kanda[1], Tomohiro Fujimoto[1],
Takuya Matsuda[1], Manik Goyal[2], Jun Yoshinobu[1], Yohei Kobayashi[1],
Takashi Oka[1], Susanne Stemmer[2], and Ryusuke Matsunaga[1]

[1]*The Institute for Solid State Physics, The University of Tokyo, Kashiwa, Chiba 277-8581, Japan*
[2]*Materials Department, University of California, Santa Barbara, California 93106-5050, USA*
*e-mail: murotani@issp.u-tokyo.ac.jp

**Abstract**

We experimentally elucidate the origin of the anomalous Hall conductivity in a three-dimensional Dirac semimetal, $Cd_3As_2$, driven by circularly polarized light. Using time-resolved terahertz Faraday rotation spectroscopy, we determine the transient Hall conductivity spectrum with special attention to its sign. Our results clearly show the dominance of direct photocurrent generation assisted by the terahertz electric field. The contribution from the Floquet-Weyl nodes is found to be minor when the driving light is in resonance with interband transitions. We develop a generally applicable classification of microscopic mechanisms of light-induced anomalous Hall conductivity.

**Main text**

Light-induced anomalous Hall effect (AHE) has attracted much interest as a potential probe of topologically nontrivial band structures tailored by light fields [1, 2]. Using the Berry curvature $\mathbf{b}(\mathbf{k})$, which quantifies the geometrical structure of wave functions in momentum space, the anomalous Hall conductivity is given by

$$\sigma_{yx} = \frac{e^2}{\hbar} \int \frac{\mathrm{d}^3 k}{(2\pi)^3} f(\mathbf{k}) b_z(\mathbf{k}), \tag{1}$$

where $f(\mathbf{k})$ is the distribution function of electrons [3, 4]. Recently, Floquet engineering using circularly polarized light (CPL) has emerged as an opportunity to change $\mathbf{b}(\mathbf{k})$ in such a manner that $\sigma_{yx}$ does not vanish. A prominent example is the Floquet-topological insulator in graphene, where a topological band gap opens at the band-touching points [5, 6]. In the case of a Dirac semimetal (DSM), a three-dimensional analogue of graphene with linearly dispersing energy bands, the degeneracy at the band-touching points is not fully lifted even under CPL; instead, they are split into pairs of doubly degenerate nodes, as shown in Fig. 1(a) [7-12]. This state is called the Floquet-Weyl semimetal, whose nontrivial topology is expected to induce a large Berry curvature around the nodes, and, concomitantly, a large contribution to $\sigma_{yx}$. This Floquet AHE promises ultrafast, reversible, and non-dissipative control of current flow in optoelectronics.

There exist, however, several competing mechanisms that could give rise to a light-induced AHE, which are physically no less important than the Floquet-Weyl semimetal. The mechanisms include the anomalous velocity of photocarriers [13, 14] and direct photocurrent generation assisted by the bias electric field [15-18]. The former arises from the light-induced change in $f(\mathbf{k})$, rather than $\mathbf{b}(\mathbf{k})$, that can also lead to a nonzero $\sigma_{yx}$. This process has important applications in the electrical detection of the spin [19, 20], valley [21, 22], and orbital [23, 24] degrees of freedom, because they can be accompanied by the Berry curvature in certain materials. Even in DSMs, CPL can excite carriers with a nonvanishing Berry curvature as shown in Fig. 1(c), which must not be neglected when interband transitions are available.

The third mechanism mentioned above, i.e., the direct photocurrent generation assisted by the bias electric field, is not included in Eq. (1). It originates from the fact that the bias field inevitably breaks the inversion symmetry of the material. Figure 1(b) shows how a bias electric field $\mathbf{E}$ along the $[\bar{1}\bar{1}1]$ axis of $Cd_3As_2$, a prototypical DSM, changes the transition probability experienced by the left-circularly polarized (LCP) light in the (112) plane. As a result of field-assisted interband transitions, excited carriers carry an intraband current,

$$\mathbf{j} = \frac{e}{\hbar} \int \frac{\mathrm{d}^3 k}{(2\pi)^3} f(\mathbf{k}) \nabla_{\mathbf{k}} \epsilon(\mathbf{k}), \qquad (2)$$

in the $[1\bar{1}0]$ direction, giving rise to a current flow perpendicular to $\mathbf{E}$. A similar mechanism has accounted for a large part of the light-induced AHE in graphene [17]. We call this phenomenon "field-induced injection current (FIIC)," by analogy with the injection current induced by the circular photogalvanic effect in noncentrosymmetric crystals [25-27]. Since the injection current is associated with the Berry curvature of energy bands through the transition probability [28, 29], the FIIC can be viewed as a manifestation of the Berry curvature engineering by a bias electric field, as opposed to an optical field as is the case for the Floquet-Weyl semimetal.

In addition to the above three mechanisms, inverse Faraday effect (IFE) and optical Kerr effect (OKE) can also contribute to the light-induced anomalous Hall conductivity. Despite the growing interest in the light-induced AHE, competition between these different mechanisms has barely been investigated. To establish the full understanding and for further exploration of the Floquet-Weyl semimetal, the ability to discriminate between these mechanisms is indispensable.

In this Letter, we resolve this complexity by terahertz (THz) Faraday rotation spectroscopy of a $Cd_3As_2$ thin film driven by CPL, as shown in Fig. 1(d). Here, a THz probe pulse serves as an ultrafast bias electric field, which initiates the anomalous Hall current in the presence of the driving light. Generation of the anomalous Hall current results in polarization rotation of the transmitted THz pulse, from which we determine the transient anomalous Hall conductivity spectrum. High temporal resolution of this technique (~100 fs) enables us to exclude extrinsic contributions to the anomalous Hall conductivity often encountered in DC measurements, since impurity scattering involved in them requires a longer time to occur [20]. Moreover, the non-contact nature of this method avoids complication by sample geometry, such as contact resistance and nonlocal response. We uncover the dominant role of FIIC based on the sign of the anomalous Hall conductivity, which has been often neglected despite its informative character. We also present general classification of the microscopic mechanisms of light-induced anomalous Hall conductivity, looking ahead to further exploration of nonlinear current generation in solids.

The sample consists of a 240 nm-thick, (112)-oriented $Cd_3As_2$ thin film, grown on a GaAs substrate with a GaSb buffer layer [30, 31]. The momentum relaxation time is as long as 190 fs, which proves the high quality of the sample. The circularly polarized pump pulse with a photon energy of 138 meV (33.3 THz in frequency, 9 μm in wavelength) selectively drives the low-energy Dirac bands [32, 33]. The other bands are left unexcited, which is important for studying the interaction between massless Dirac fermions and light. All experiments are performed at room temperature.

Figure 2(a) shows the pump-induced change in the polarization component parallel to the incident THz probe, $\Delta E_x$, as a function of the probe delay (horizontal axis) and the pump delay (vertical axis). We found no dependence on the pump helicity, which led us to present the average for the LCP and right-circularly polarized (RCP) pump pulses. The signal seen here corresponds to a decrease in the transmittance. The change in the longitudinal conductivity $\sigma_{xx}$ obtained from these data is shown in Fig. 2(b) as a function of frequency (horizontal axis) and the pump delay (vertical axis). The spectrum exhibits an overall increase, and finally approaches a Drude-type response. This behavior reflects intraband absorption by photoexcited carriers [34].

Figure 2(c) shows the pump-induced change in the polarization component perpendicular to the incident probe, $\Delta E_y$. Here, half the difference between the results for the LCP and RCP pump pulses is presented, so as to extract the helicity-dependent signal caused by the light-induced anomalous Hall conductivity. One can see clear polarization rotation around the pump irradiation time (pump delay $\Delta t \simeq 0$ ps). We determine the transient Hall conductivity according to

$$\Delta \sigma_{yx}(\omega) = -\frac{1 + n_{\text{subs}} + \sigma_{xx}(\omega) Z_0 d}{Z_0 d} \Delta \theta(\omega), \qquad (3)$$

where $n_{\text{subs}}$ is the refractive index of the substrate, $d = 240$ nm the thickness of the sample, $Z_0 = 377$ Ω the vacuum impedance, and $\Delta \theta(\omega) = \Delta E_y(\omega)/E_x(\omega)$ the complex polarization rotation angle [35]. Figure 2(d) shows the real part of $\Delta \sigma_{yx}(\omega)$ as a function of frequency (horizontal axis) and the pump delay (vertical axis). It is clear that an LCP pump pulse induces a positive Hall conductivity in the entire frequency region at $\Delta t \simeq 0$ ps. Figure 3(a) traces the temporal change of the anomalous Hall conductivity at 6.2 meV. The long-lasting signal after 0.5 ps arises from the extrinsic contributions to the anomalous velocity of photocarriers, which will be discussed elsewhere.

As shown in Supplemental Material [36], the Floquet-Weyl semimetal induces a negative anomalous Hall conductivity for LCP pump pulses,

$$\sigma_{yx} = -\frac{N_\mathrm{D} e^4 v E_\mathrm{pump}^2}{2\pi^2 \hbar^3 \Omega^3}, \quad (4)$$

where $N_\mathrm{D}$ is the number of Dirac nodes, $v$ the Fermi velocity, $E_\mathrm{pump}$ the amplitude of the pump pulse, and $\Omega$ its frequency. The negative sign contradicts the experimental result in Fig. 3(a), so we can exclude the Floquet-Weyl semimetal from being the primary origin of the anomalous Hall conductivity. Anomalous velocity of photocarriers also fails to explain the positive sign observed in experiment; as shown in Fig. 1(c), an LCP pump excites electron-hole pairs with a negative Berry curvature $b_{[221]} < 0$, which leads to a negative anomalous Hall conductivity $\sigma_{yx} < 0$ according to Eq. (1). By contrast, FIIC induces a positive Hall conductivity for an LCP pump. Defining the $[\bar{1}\bar{1}1]$, $[1\bar{1}0]$, and $[221]$ directions in Fig. 1(b) as $x$, $y$, and $z$ axes, one can see that a Hall current $j_y > 0$ is generated by the LCP light propagating in the $+z$ direction in the presence of $E_x > 0$, giving rise to a positive Hall conductivity $\sigma_{yx} = j_y/E_x > 0$. As long as the two-band description is valid, the sign of the anomalous Hall conductivity in each mechanism does not depend on details of the band structure or orientation of the sample, which makes it a reliable indicator of the microscopic origin [36]. Remarkably, the observed anomalous Hall conductivity has not only a sign opposite to but also a magnitude larger than the theoretical prediction for a Floquet-Weyl semimetal. Figure 3(b) shows the fluence dependence of $\mathrm{Re}\,\Delta\sigma_{yx}$ at 6.2 meV. In the weak excitation limit ($< 20$ μJ/cm$^2$), it is proportional to the pump intensity, being consistent with the FIIC originating from the interband transitions assisted by the THz electric field. The observed conductivity is larger than the estimation by Eq. (4) for the Floquet-Weyl semimetal (dashed line), which is consistent with microscopic theory [36]. These observations provide strong evidence for the dominant role of FIIC.

As another possible origin of the light-induced anomalous Hall conductivity, we mention the IFE. The IFE arises when CPL generates a net magnetization, which can cause THz Faraday rotation in a manner similar to ferromagnets. In fact, contribution from the magnetization by itinerant carriers have already described as the anomalous velocity of photocarriers. Magnetization by localized spins is negligible in the present case, because it arises from nonresonant virtual transitions much weaker than the resonantly excited ones [37]. OKE via higher-energy transitions can also be neglected because of its nonresonant character.

Let us discuss the origin of the temporal oscillation seen in Fig. 3(a). Equation (3) assumes that the electric field emitted by a current density **j** follows

$$\mathbf{E}_{\text{em}}(\omega) = -\frac{Z_0 d}{1 + n_{\text{subs}}} \mathbf{j}(\omega), \tag{5}$$

which is valid for a thin and planer source. To be exact, this assumption fails at low frequencies, where the transverse size of the source becomes smaller than the wavelength. The emitted field then diverges so much that a part of it escapes from the parabolic mirror collecting the transmitted wave (see Fig. S6(b) in Supplemental Material [36]). Moreover, upon focusing onto the detector by another parabolic mirror, worse diffraction limit at longer wavelengths lowers the detection efficiency. As a result, the low-frequency components of the emitted wave are filtered out before detection, leading to an oscillatory waveform. Being a consequence of propagation, this oscillation appears along the real time, $t$. In Fig. 2(c), the $t$ axis corresponds to the (1, 1) direction, because the horizontal and vertical axes of this figure are defined as $X = t - t_{\text{probe}}$ and $Y = t - t_{\text{pump}}$, respectively, where $t_{\text{probe}}$ and $t_{\text{pump}}$ denote the arrival times of each pulse; motion on a diagonal line ($Y = X + \text{const.}$) corresponds to passage of $t$ with a fixed time difference between the pump and probe ($t_{\text{probe}} - t_{\text{pump}}$). The temporal oscillation thus extends in the (1, 1) direction in Fig. 2(c), and is projected onto the one-dimensional cut plotted in Fig. 3(a), which accounts for the oscillation in the latter. Despite such complication by the frequency filtering, the high-frequency data remains reliable, including the sign of the measured anomalous Hall conductivity. To corroborate our interpretation, we perform a model calculation of the emission by the anomalous Hall current, taking the frequency filtering effect into account. The result is presented in Fig. 3(d), showing excellent agreement with the experimental result in Fig. 2(c) including the sign of $\Delta E_y$. Details of the simulation are given in Supplemental Material [36].

To get a more comprehensive view on the light-induced anomalous Hall conductivity, we classify the examined mechanisms on the basis of normal and anomalous velocities, which can also be extended to different systems. Velocity of Bloch electrons is generally given by

$$\mathbf{v}(\mathbf{k}) = \frac{1}{\hbar} \nabla_{\mathbf{k}} \epsilon(\mathbf{k}) - \frac{e}{\hbar} \mathbf{E} \times \mathbf{b}(\mathbf{k}), \tag{6}$$

where the first term corresponds to the normal velocity, while the second term represents the anomalous velocity induced by the Berry curvature $\mathbf{b}(\mathbf{k})$ and an electric field $\mathbf{E}$ [3, 4]. The total electric current, $\mathbf{j} = e \sum_{\mathbf{k}} f(\mathbf{k}) \mathbf{v}(\mathbf{k})$, is determined by the velocity $\mathbf{v}(\mathbf{k})$ as well as the electron distribution function $f(\mathbf{k})$. Because the light-induced anomalous Hall conductivity originates from a third-order nonlinearity, we focus on currents proportional to the product of the pump intensity $I \propto |E_0|^2$ and the probe electric field $E_x$. The relevant third-order current is given by

$$\mathbf{j}^{(3)} = e \sum_{\mathbf{k}} (f_0 \mathbf{v}_3 + f_1 \mathbf{v}_2 + f_2 \mathbf{v}_1 + f_3 \mathbf{v}_0). \tag{7}$$

The subscript denotes the order with respect to the electric field, as explicitly given below. In the first term, $f_0$ denotes the equilibrium distribution independent of any electric fields, while $\mathbf{v}_3 \propto$

$E_x I$ represents the anomalous velocity arising from the pump-induced Berry curvature. This term includes the contribution by the Floquet-Weyl semimetal. For this mechanism to overcome the other terms, the photoexcited carrier density must be negligibly small. This is an important issue, because Floquet engineering requires strong light fields which inevitably excite carriers unless the material is transparent. In the second term, $f_1 \propto E_x$ represents the shift in the distribution function caused by the probe field, while $\mathbf{v}_2 \propto I$ arises from a pump-induced change in the dispersion relation. We consider such a process to be of minor importance. In the third term, $f_2 \propto I$ is induced by ordinary one-photon absorption, while $\mathbf{v}_1 \propto E_x$ is nothing but the anomalous velocity. This term thus corresponds to the anomalous velocity of photocarriers. In the fourth term, $f_3 \propto E_x I$ is induced either by interband transitions assisted by the probe field, or by intraband acceleration of photoexcited carriers, while $\mathbf{v}_0$ is the normal velocity. The FIIC belongs to this term. We summarize the mechanisms relevant to Cd$_3$As$_2$ in Table I. Interband transitions are essential when the driving light is resonant to them, which is a natural situation for gapless DSMs.

Finally, we revisit the intensity dependence from a viewpoint of Floquet states. As the pump intensity exceeds 20 μJ/cm$^2$, increase of the anomalous Hall conductivity is slowed down [Fig. 3(b)], which is naively attributed to the Pauli blocking by excited carriers. However, saturation of the anomalous Hall conductivity does not faithfully trace the suppression of the simple one-photon absorption shown in Fig. 3(c). This discrepancy may indicate the effect of Floquet state formation at high excitation intensities. A similar intensity dependence of the anomalous Hall conductivity has been observed and calculated for graphene, where the participation of light-dressed Floquet states has been established [6, 17, 38]. Roles of the Floquet states in the presence of complicated scattering channels are still under intensive investigation [17, 38-43]. Even more interestingly, saturation of the FIIC may enable dominance of the Floquet-Weyl semimetal at much higher excitation intensities. We expect the Floquet-Weyl semimetal to be more robust against saturation, because it is not affected by Pauli blocking by the photoexcited carriers. To explore this possibility, the Fermi level, which lies ~50 meV above the Dirac nodes at present, should be brought closer to the nodes, because doped carriers may reduce the anomalous Hall conductivity. A combination of chemical and electrical doping enables the Fermi level to be tuned in Cd$_3$As$_2$ [44], which is promising in this direction.

In summary, we experimentally studied the light-induced anomalous Hall conductivity in a 3D DSM. Taking advantage of time-resolved THz Faraday rotation spectroscopy, we unambiguously identified the FIIC as the dominant origin of the anomalous Hall conductivity during irradiation by CPL. Our observation paves the way for ultrafast Berry curvature engineering with THz pulses, which activate the circular photogalvanic effect in a highly controllable manner. We also find that transparency to the driving light is the key to detect the Floquet-Weyl semimetal through the AHE. Our experimental technique will be a powerful tool to unveil dynamical aspect of the light-induced AHE, with the complex interplay of competing mechanisms disentangled.

**Acknowledgments**

This work was supported by JST PRESTO (Grant Nos. JPMJPR20LA and JPMJPR2006), JST CREST (Grant No. JPMJCR20R4), and in part by JSPS KAKENHI (Grants Nos. JP19H01817, JP20J01422, and JP20H00343). RM also acknowledges partial support by Attosecond lasers for next frontiers in science and technology (ATTO) in Quantum Leap Flagship Program (MEXT Q-LEAP). S.S. and M.G. acknowledge support by CATS Energy Frontier Research Center, which is funded by the Department of Energy, Basic Energy Sciences, under contract DE-AC02-07CH11358.

R.M. conceived the project. M.G. fabricated the sample with guidance from S.S. N.K. and T.M. evaluated the linear response function. Y.M., N.K., and T.F. developed the pump-probe spectroscopy system with the help of J.Y., Y.K., and R.M. Y.M. performed the pump-probe experiment and analyzed the data with N.K. Y.M. conducted the theoretical calculations with the help of T.O. All the authors discussed the results. Y.M. prepared the manuscript with substantial feedbacks from R.M., T.O., S.S., and all the coauthors.

**Figures and figure captions**

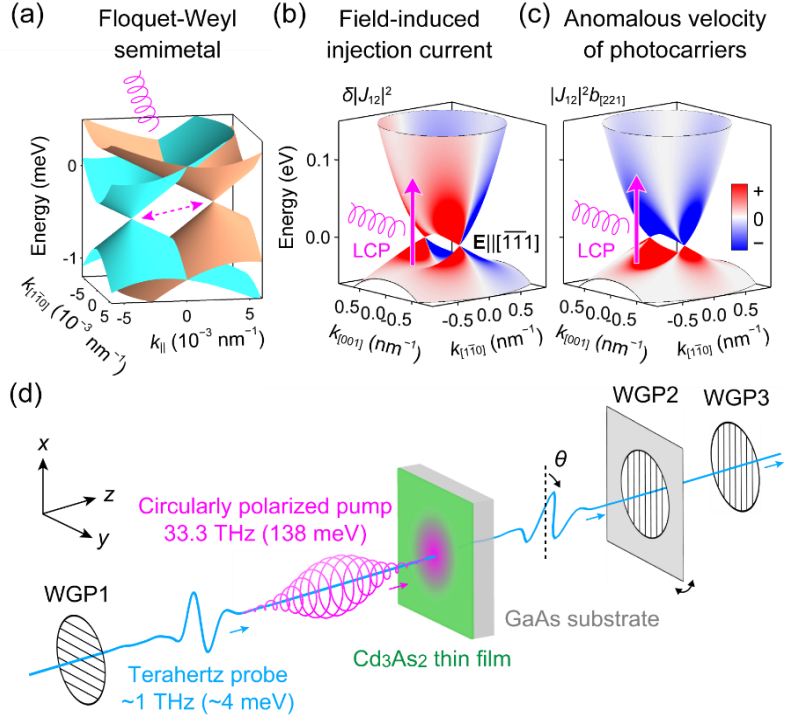

FIG. 1. (a) Energy dispersion relation of the Floquet Weyl semimetal in $Cd_3As_2$ for CPL normally incident on the (112) plane. $k_{[1\bar{1}0]}$ measures the electron momentum in the $[1\bar{1}0]$ direction with respect to the center of the Floquet-Weyl nodes. $k_{\parallel}$ denotes the momentum in the direction of splitting, which is perpendicular to $k_{[1\bar{1}0]}$. (b) Change in the transition probability, $\delta|J_{12}|^2$, experienced by the LCP light propagating in the [221] direction of $Cd_3As_2$, caused by a bias electric field $\mathbf{E}||[\bar{1}\bar{1}1]$. (c) Product of the transition probability $|J_{12}|^2$ and the Berry curvature $b_{[221]}$ of electrons, for the LCP light propagating in the [221] direction of $Cd_3As_2$. (d) Setup of the pump-probe experiment. A rotatable wire grid polarizer (WGP) determines the polarization state of the transmitted THz probe pulse.

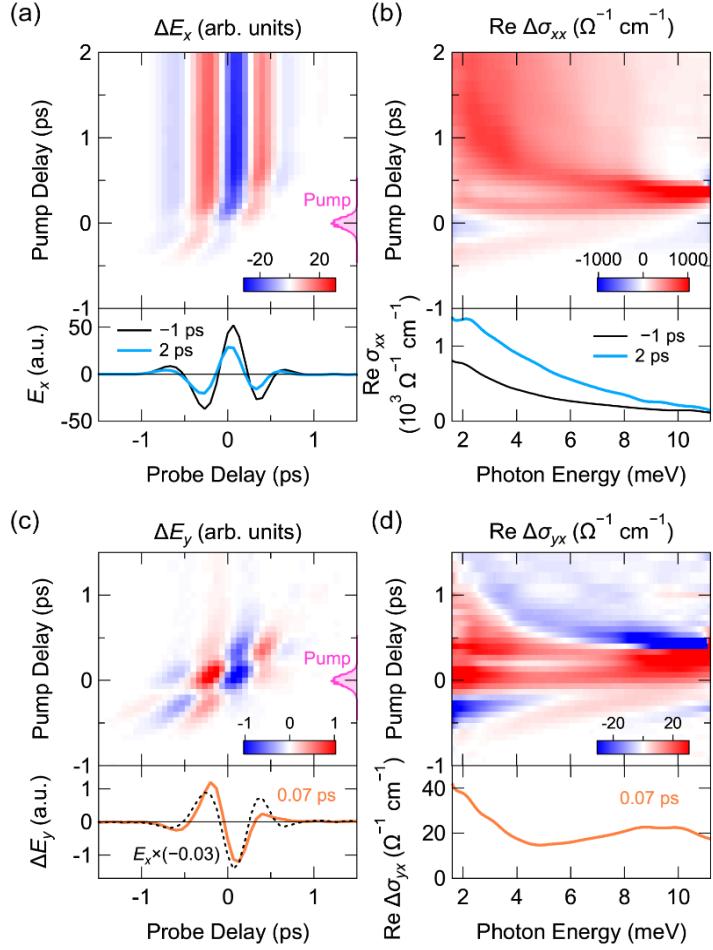

FIG. 2. (a) Top: Pump-induced change in the $x$-component of the transmitted THz probe, $\Delta E_x$, as a function of the probe delay (horizontal axis) and the pump delay $\Delta t$ (vertical axis). The data is shown for $(\text{LCP pump} + \text{RCP pump})/2$ with a pump fluence of 132 μJ/cm$^2$, corresponding to a peak electric field of 89 kV/cm inside the sample. The pump intensity is overlaid on the right side. Bottom: Waveforms of $E_x$ at $\Delta t = -1$ ps (thin black line) and 2 ps (thick cyan line). a.u. stands for arbitrary units. (b) Top: Change in the real part of the longitudinal conductivity, $\text{Re}\,\Delta\sigma_{xx}$, as a function of frequency (horizontal axis) and the pump delay $\Delta t$ (vertical axis). Bottom: Longitudinal conductivity ($\text{Re}\,\sigma_{xx}$) at $\Delta t = -1$ and 2 ps. (c) Top: Pump-induced change in the $y$-component of the transmitted THz probe, $\Delta E_y$. The data is shown for $(\text{LCP pump} - \text{RCP pump})/2$. Bottom: Waveform of $\Delta E_y$ at $\Delta t = 0.07$ ps (solid line), along with the corresponding waveform of $E_x$ multiplied by $(-0.03)$. (d) Top: Change in the real part of the Hall conductivity, $\text{Re}\,\Delta\sigma_{yx}$. Bottom: Hall conductivity spectrum at $\Delta t = 0.07$ ps.

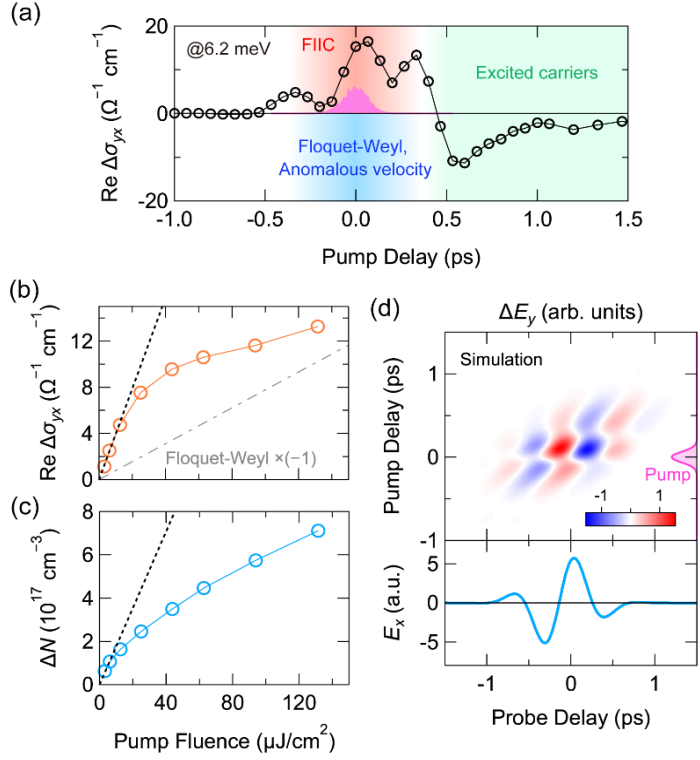

FIG. 3. (a) Pump delay dependence of the anomalous Hall conductivity at 6.2 meV (circle), extracted from the data in Fig. 2(d). A positive signal in the pump-probe overlap ($\Delta t \simeq 0$ ps) indicates the dominance of the FIIC, while a negative one would indicate the Floquet-Weyl semimetal or the anomalous velocity of photocarriers. The shaded curve shows the pump intensity. (b) Fluence dependence of the maximum value in the pump delay-dependent anomalous Hall conductivity at 6.2 meV (circle). The dotted line shows a linear fit to the low-fluence data. The dashed line gives a theoretical prediction for the Floquet-Weyl semimetal multiplied by $-1$. (c) Density of excited carriers obtained from a Drude model fitting to the longitudinal conductivity $\sigma_{xx}$ at the pump delay $\Delta t = 2$ ps. The dotted line shows a linear fit to the low-fluence data. (d) Top: Simulated emission $\Delta E_y$ from the transient Hall current. Bottom: Waveform of the incident THz probe.

**Tables**

TABLE I. Origin of anomalous Hall conductivity under circularly polarized light field.

| Mechanism | Floquet-Weyl semimetal | Field-induced injection current | Anomalous velocity of photocarriers |
|---|---|---|---|
| Driving force | Change of Berry curvature | Population with momentum | Population with Berry curvature |
| Terms in Eq. (7) | 1st | 4th | 3rd |
| $\sigma_{yx}$ For LCP | − | + | − |
| This experiment | Minor | Dominant | Minor |